\documentclass[10pt, conference, letterpaper]{IEEEtran}

\IEEEoverridecommandlockouts

\usepackage{cite}
\usepackage{amsmath,amssymb,amsfonts}
\usepackage{algorithmic}
\usepackage{graphicx}
\usepackage{textcomp}
\usepackage{xcolor}
\usepackage{url}
\def\BibTeX{{\rm B\kern-.05em{\sc i\kern-.025em b}\kern-.08em
    T\kern-.1667em\lower.7ex\hbox{E}\kern-.125emX}}

\begin{document}

\title{ONDT: A Modular and Virtualized O-RAN Digital Twin for xApp/rApp Closed-Loop Experimentation\\
\thanks{This work is supported by the European Union’s Horizon Europe research through SUNRISE-6G (GA No. 101139257) and SOVEREIGN-6G (GA No. 101292787) projects. It also received support from the COALESCE-6G project (PID2024-163028OB-I00), funded by MICIU/AEI/10.13039/501100011033/FEDER, EU. i2CAT authors further acknowledge the support of the CERCA Programme of the Generalitat de Catalunya.}
}
\author{\IEEEauthorblockN{Miguel Catalan-Cid,\\Joan Josep Aleixendri,\\August Betzler}
        \IEEEauthorblockA{
            \textit{i2CAT Foundation}\\
            \{miguel.catalan,  joan.aleixendri,august.betzler\}@i2cat.net 
        }
        \and
\IEEEauthorblockN{George Oikonomou,\\Lydia Ntzogani}
\IEEEauthorblockA{
            \textit{Juniper Networks}\\
            \{goikonomou,\\ntzoganil\}@juniper.net
        }
                \and
\IEEEauthorblockN{David Reiss}
\IEEEauthorblockA{
            \textit{UPC, ETSETB}\\
            \{david.reiss\}@upc.edu 
        }
}

\maketitle

\begin{abstract}
Open Radio Access Network (O-RAN) architectures extend programmability in next-generation cellular networks through Near-Real-Time and Non-Real-Time RAN Intelligent Controllers (RICs), enabling AI/ML-driven xApps and rApps to optimize network behavior dynamically. However, validating automated closed-loop control algorithms in live environments presents significant operational risks and deployment costs. To address this challenge, this paper introduces the O-RAN Digital Twin (ONDT), a modular, virtualized, and cloud-native experimentation framework developed as a 6G enabler within the SNS SUNRISE-6G project. We present the ONDT architecture and its end-to-end experimentation workflows, showing how the framework automates O-RAN-compliant rApp/xApp onboarding, RAN scenario configuration and initialization, Key Performance Metric (KPM) collection, and closed-loop control validation. Experimental validation using Juniper RIC platforms and the Keysight RICtest emulation framework demonstrates ONDT’s ability to provide an isolated and reproducible environment for KPM monitoring, cell ON/OFF control, automated dataset generation, and rApp/xApp validation.
\end{abstract}


\section{Introduction and Motivation}\label{sec:intro}

Open Radio Access Network (O-RAN) is reshaping how Radio Access Networks (RANs) are managed. By introducing programmable control through RAN Intelligent Controllers (RICs), O-RAN enables a software-driven ecosystem where third-party applications can monitor network conditions and influence RAN behavior through standardized interfaces and data models. In this architecture, the Near-Real-Time (RT) RIC hosts xApps that implement near-RT control loops using E2-based telemetry and control procedures, while the Non-Real-Time RIC hosts rApps that support longer-term analytics, A1-based policy management, AI/ML-assisted optimization, and RAN configuration support through the SMO framework and the O1 interface~\cite{understanding_oran}. This split enables flexible closed-loop control, but it also makes validation more complex: applications must be assessed not only as algorithms, but as deployed software components interacting with RIC platforms, service models, open interfaces, telemetry pipelines, and RAN behavior.

This validation challenge becomes more relevant as the O-RAN application ecosystem matures. Vendor platforms are already moving toward dedicated environments for rApp development, testing, and deployment. Ericsson’s Intelligent Automation Platform (EIAP) exposes a Non-RT RIC-based automation platform and an SDK for developing and deploying rApps, while the EIAP Ecosystem supports the build and validation of automation applications through dedicated test and validation environments~\cite{EricssonEIAP2026}. Nokia follows a similar direction with its MantaRay SMO portfolio, which supports an rApp ecosystem involving Nokia and application developers~\cite{NokiaMantaRaySMO2026}. Beyond vendor-specific platforms, BubbleRAN’s TelcoFabric initiative illustrates the emergence of shared repositories for reusable telco intelligence artifacts, including xApps, rApps, datasets, models, and benchmarks~\cite{telcofabric}. These developments indicate that rApps and xApps are becoming part of broader multi-party software ecosystems. As a result, repeatable validation workflows are required before such applications are introduced into operational RAN environments.

Conventional testing methods only partially address this need. Offline evaluation over datasets can support algorithm design, training, calibration, and benchmarking, particularly when the data reflects realistic traffic and RAN conditions \cite{practical}. Work on commercial-traffic twinning has shown that reproducing realistic traffic traces on experimental Open RAN platforms can improve the relevance of RIC-application evaluation~\cite{twinning}. Similarly, physical O-RAN experimentation platforms have collected real measurements from virtualized base-station deployments, including datasets for computing-resource usage, energy consumption, and edge-application performance under different radio, traffic, and computing configurations~\cite{dataset_nec}.  However, these approaches are not sufficient to validate complete O-RAN application workflows~\cite{testbed}. Recent multi-vendor Open RAN evaluations show that, even when individual interface procedures can be verified, end-to-end integration still requires substantial alignment across RAN components, RIC implementations, service models, configuration parameters, and software versions~\cite{interoperability}.

In practice, xApps and rApps must be onboarded, instantiated, configured, connected to the appropriate RIC services, exposed to representative RAN measurements, and evaluated under repeatable network conditions. Moreover, when rApps and xApps cooperate within the same control loop, the resulting behavior depends on the interaction between Non-RT and Near-RT logic and on the response of the underlying RAN. Recent work on Tandem Apps has highlighted this type of coordinated rApp/xApp design, where longer-term global decisions and faster local actions are split across the two RIC layers~\cite{rimedo}. Such architectures further reinforce the need for integrated testing environments able to exercise the full cross-RIC control loop, rather than validating each application component in isolation.

Digital Twins (DTs) have emerged as a promising mechanism to support this intermediate validation stage between offline analysis and live-network deployment. In the O-RAN context, a DT can provide a controlled virtual environment where RAN scenarios, traffic conditions, telemetry streams, RIC applications, and control actions can be evaluated before being applied to operational infrastructure. Recent O-RAN  studies have identified DT-RAN as a relevant direction for AI/ML training, evaluation, performance assurance, and network testing automation~\cite{orandt}. These studies emphasize the need for reproducible environments where RICs, xApps, rApps, RAN components, and multi-vendor interfaces can be tested through automated pipelines.

Several recent works have started to explore O-RAN DT architectures, implementations, and complementary experimentation platforms. Colosseum provides a high-fidelity Open RAN DT based on hardware-in-the-loop RF emulation and end-to-end softwarized O-RAN and 5G protocol stacks,  enabling repeatable testing and integration of Open RAN control applications~\cite{colosseum}. Open-source approaches have also connected network simulators with Near-RT RIC platforms and O-RAN-compliant xApps to validate use cases such as load balancing and energy efficiency~\cite{dt_eucnc}, while broader architectural studies have proposed O-RAN DT concepts based on modular building blocks, including traffic models, RAN models, RF/mobility models, scenario generators, and integration with RIC functions~\cite{dt_commag}.

In this context, this paper presents the O-RAN Digital Twin (ONDT), a modular, virtualized, and cloud-native experimentation framework developed within the SNS SUNRISE-6G project~\footnote{https://sunrise6g.eu/}. ONDT supports controlled validation of O-RAN applications and closed-loop workflows before deployment by translating high-level experiment requests into operations over the RIC and emulated RAN domains. The paper describes the ONDT architecture and the main workflows for rApp/xApp experimentation, including application onboarding, scenario execution, KPM collection, dataset generation, and control-action validation. ONDT is validated using Juniper RIC and Keysight RICtest platforms, demonstrating xApp-based KPM monitoring, rApp-assisted dataset generation, and joint rApp/xApp closed-loop validation.

The remainder of the paper is organized as follows. Section~\ref{sec:arch} presents the ONDT architecture and its main components. Section~\ref{sec:workflows} describes the experimentation workflows supported by the framework. Section~\ref{sec:validation} details the validation environment and functional tests. Finally, Section~\ref{sec:conclusion} concludes the paper and outlines future work.

\section{ONDT Architecture}\label{sec:arch}

The ONDT is implemented in SUNRISE-6G as a 6G Enabler platform for controlled experimentation, dataset generation, and pre-deployment testing of xApps and rApps. The ONDT focuses on three O-RAN domains: the Near-RT RIC, the Non-RT RIC, and the RAN domain, which is emulated by the simulation layer. 

\begin{figure}
    \centering
    \includegraphics[width=1\linewidth]{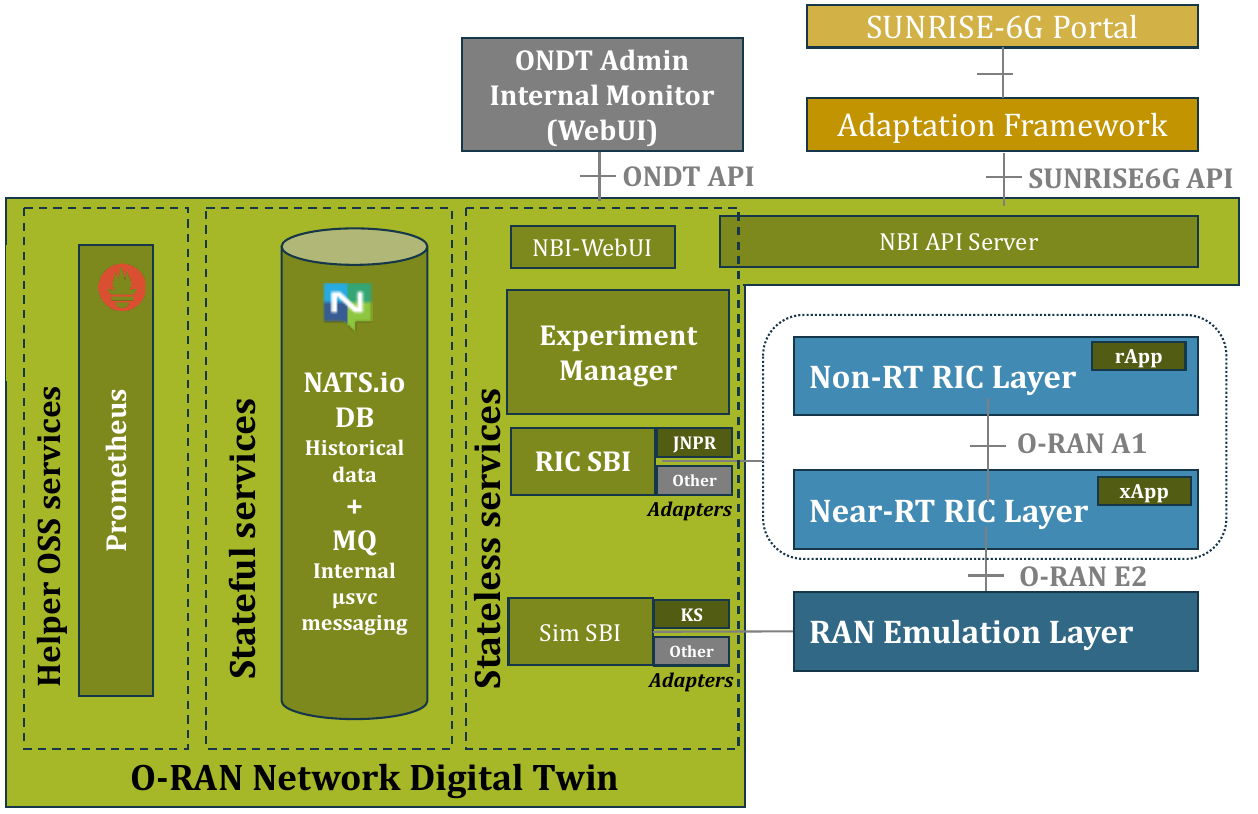}
    \caption{High-level ONDT architecture with Juniper and Keysight adapters for the RIC and RAN emulation layer.}
    \label{fig:fig1_arch}
\end{figure}

Figure~\ref{fig:fig1_arch} shows the high-level ONDT architecture. The ONDT layer is placed between the SUNRISE-6G Adaptation Framework, which is connected to the SUNRISE-6G Portal, and the RICs and RAN emulation layer. Its main role is to translate high-level experiment requests into low-level vendor-specific commands required to execute an ONDT experiment. The ONDT framework is implemented as a set of Kubernetes-orchestrated microservices. All services are stateless except the message queue, which is based on NATS and is used both for internal asynchronous communication and for storing configuration data, experiment history, runtime status, and other ONDT state information.

The northbound part of the ONDT consists of the NBI API Server and the NBI WebUI. The NBI API Server exposes the service consumed by the SUNRISE-6G Adaptation Framework. Generated from the Adaptation Framework OpenAPI specifications, it validates incoming HTTP requests, and publishes valid requests to the NATS message queue. The NBI WebUI provides read-only monitoring for the ONDT administrator, including the status of experiments, ONDT components, RIC components, RAN emulator components, and deployed applications.

The Experiment Manager is the central ONDT component. It handles the experiment lifecycle, stores configuration and state, and tracks the real-time status of the southbound components. Through the southbound interfaces, it can trigger rApp/xApp onboarding and deployment, load and initialize a network topology in the RAN emulator, start and stop experiments, and execute offboarding operations after experiment completion or forced termination.

The SBI RIC component provides the interface toward the Non-RT and Near-RT RICs. Internally, it is designed to be independent of the specific RIC implementation, while externally it translates ONDT requests into the proprietary API calls required by the target RIC platform. It supports adding and removing RICs, onboarding and offboarding rApps and xApps, deploying and undeploying applications, and monitoring RIC operations. Within the scope of the SUNRISE-6G project, this component was tailored to Juniper Non-RT and Near-RT RICs~\footnote{https://www.juniper.net/gb/en/products/network-automation/ran-intelligent-controller.html}. Juniper's Non-RT RIC includes an rApp SDK, A1 termination, and R1 termination, while the Near-RT RIC includes an xApp SDK, an A1 mediator, and E2 termination. Both RICs also include an Application On-boarder and an Application Manager to control the lifecycle of rApps and xApps. These components support the validated workflows in which rApps obtain RAN telemetry, generate A1 policies according to data- and AI-driven control logic, and xApps process those policies, perform E2-based KPM reporting, and issue E2-based RAN control or configuration actions.

The SBI RAN emulator component provides the equivalent interface toward the RAN emulation layer. It supports listing available network topologies, loading a topology into the emulator, starting a topology or experiment so that E2 messages flow toward the Near-RT RIC, stopping a running topology, and collecting results. Within the scope of SUNRISE-6G, it was integrated and validated with Keysight RICtest~\footnote{https://www.keysight.com/sg/en/product/P8828S/rictest-ran-intelligent-controller-test-solutions.html}. This solution supports Key Performance Measurement (KPM) RAN Control (RC), and Cell Configuration and Control (CCC) E2 Service models, enabling telemetry gathering, network-state reporting, and RAN-state modification. It also allows the configuration of sites, capacity and coverage cells, cell parameters, slices, radio resources per slice, UE mobility, traffic rates, and traffic timelines through JSON-based scenario descriptions. 

The ONDT also includes basic monitoring and data-collection support.  Prometheus is used so that rApps and xApps can push or retrieve runtime metrics during experiment execution. The collected metrics can be used for monitoring and dataset generation. 

Through these components, ONDT supports the main experiment workflows required in SUNRISE-6G: experiment selection from the Portal, rApp/xApp onboarding, topology loading, KPM collection, dataset generation, data- and AI-driven closed-loop execution, A1 policy delivery, and E2-based control through RC and CCC messages.

\section{Experimentation Workflows}\label{sec:workflows}

The ONDT architecture supports the execution of O-RAN experiments through a set of reusable workflows that coordinate the SUNRISE-6G Portal, the ONDT orchestration layer, the RIC platforms, and the RAN emulation layer. These workflows are designed to be independent of a specific use case: once the required rApps, xApps, and RAN scenarios are available, the ONDT can select, deploy, execute, monitor, and terminate experiments through the same lifecycle.

\subsection{Baseline Experiment Lifecycle}

\begin{figure}
    \centering
    \includegraphics[width=1\linewidth]{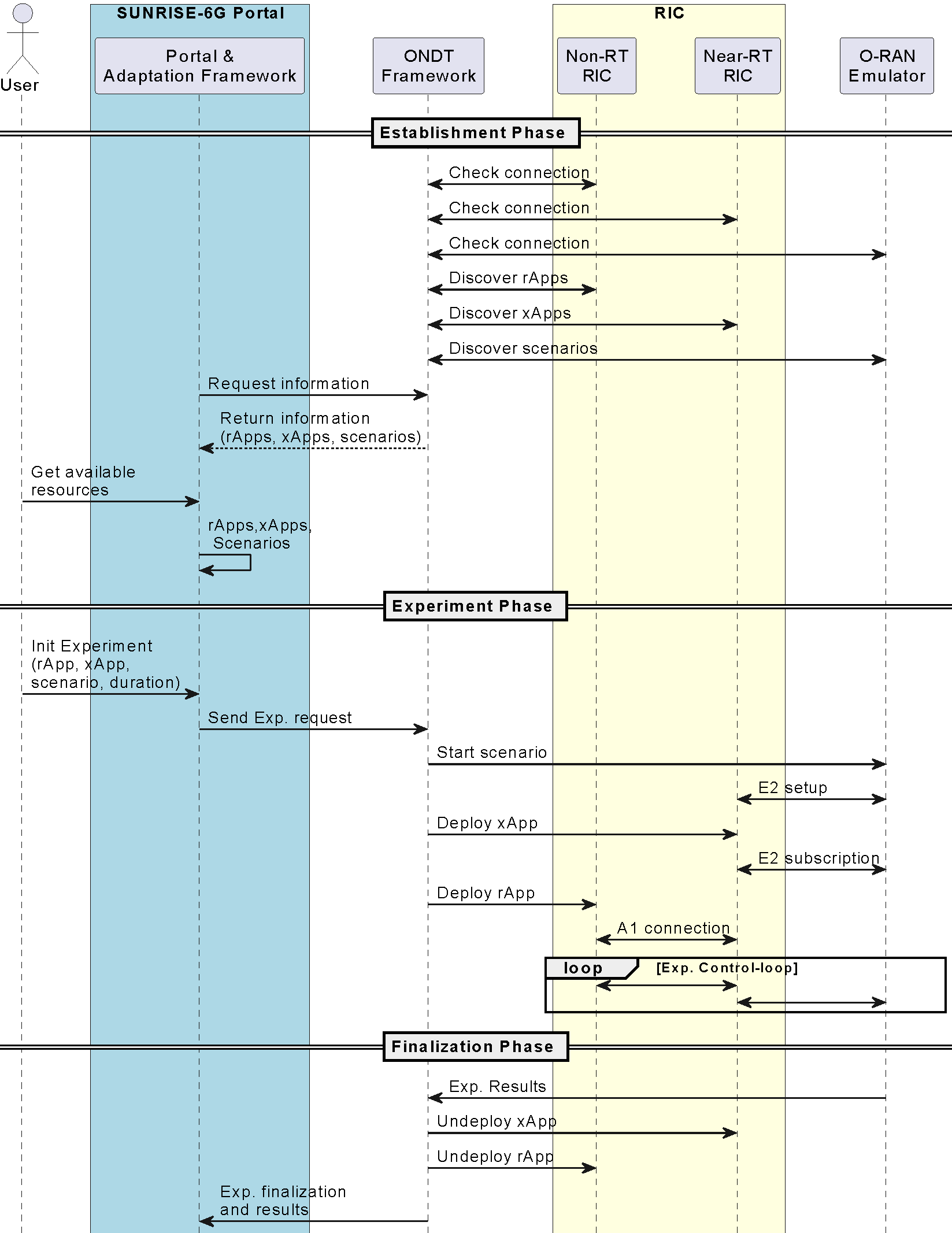}
    \caption{Baseline ONDT experimentation workflow.}
    \label{fig:basic_workflow}
\end{figure}

Figure~\ref{fig:basic_workflow} illustrates the baseline ONDT experimentation workflow, which is divided into establishment, experiment, and finalization phases. During the establishment phase, ONDT checks connectivity with the Non-RT RIC, the Near-RT RIC, and the O-RAN emulator. It also discovers the available rApps, xApps, and emulated scenarios, and exposes this information to the SUNRISE-6G Portal so that the experimenter can select the resources required for an experiment.

During the experiment phase, the user initializes an experiment by selecting the target rApp, xApp, scenario, and duration. The SUNRISE-6G Portal forwards the experiment request to the ONDT, which coordinates the required southbound operations. These operations include starting the selected scenario in the O-RAN emulator, deploying the selected xApp in the Near-RT RIC, and deploying the selected rApp in the Non-RT RIC. The E2 setup between the emulator and the Near-RT RIC is established when the scenario is started, while A1 connectivity is established once the corresponding rApp and xApp are instantiated. These interactions enable the execution of the experiment control loop.

During the finalization phase, the experiment results generated by the RAN emulation layer are retrieved through the Simulation SBI, stored in the NATS.io database, and forwarded to the SUNRISE-6G Portal, where they become available after experiment finalization. Finally, the ONDT undeploys the rApp and xApp, completing the experiment lifecycle.

\subsection{KPM Monitoring and Dataset Generation}

\begin{figure}
    \centering
    \includegraphics[width=1\linewidth]{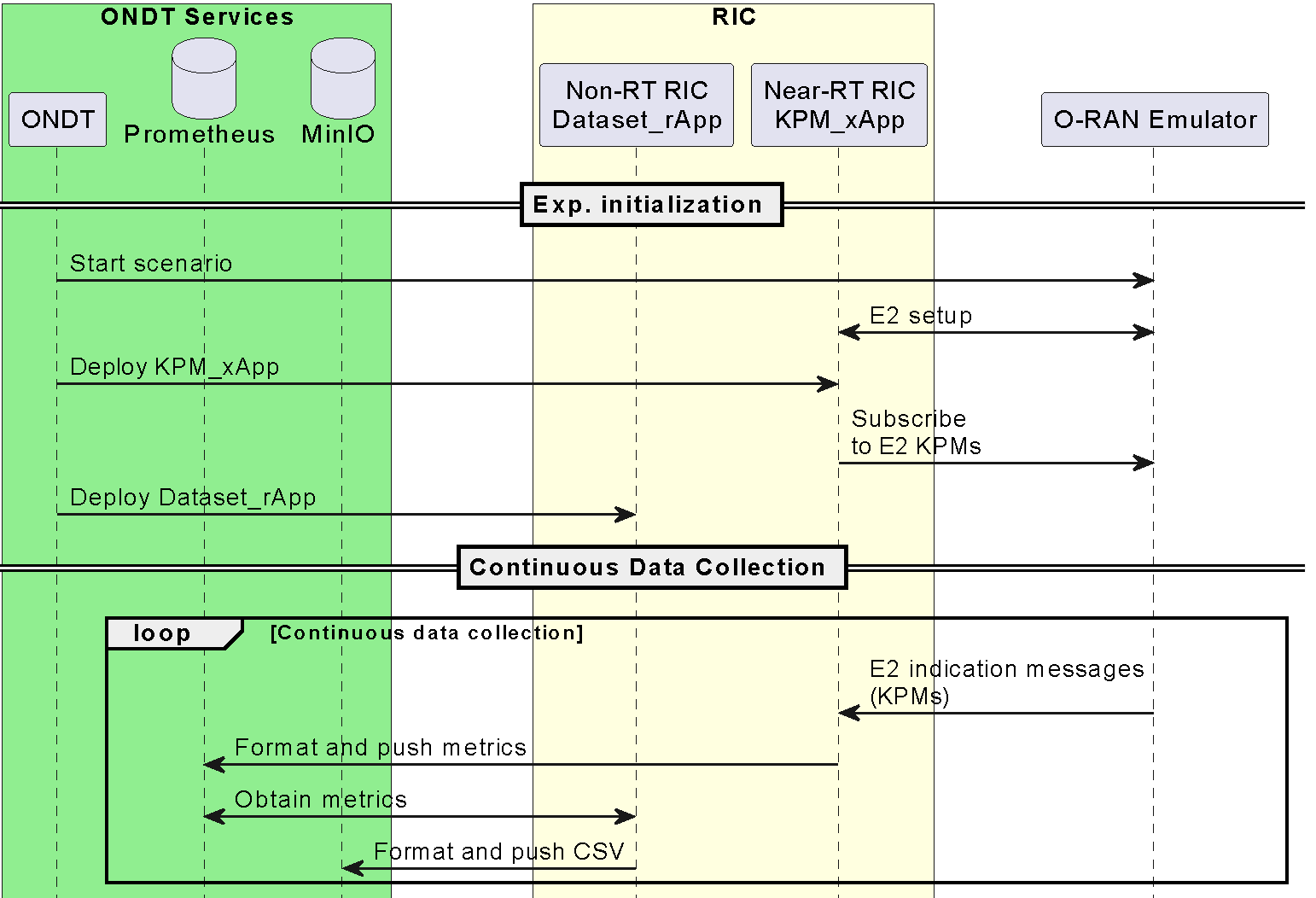}
    \caption{KPM Monitoring and Dataset Generation workflow.}
    \label{fig:kpm_workflow}
\end{figure}

The second basic ONDT workflow focuses on KPM monitoring and dataset generation, illustrated in Figure~\ref{fig:kpm_workflow}. While the baseline workflow retrieves the raw results generated by the RAN emulation layer, this workflow enables experiment-specific telemetry collection through the E2-KPM service model. In this case, the selected xApp subscribes to the required E2 nodes, cells, and KPMs, allowing the experimenter or xApp developer to collect the measurements needed by the target use case.

After the selected RAN scenario is initialized and the E2 setup is completed, the xApp retrieves the information required to identify the available E2 nodes and to generate the corresponding RIC Subscription Request. In the implemented ONDT validation, the subscription follows \texttt{E2SM-KPM-EventTriggerDefinition-Format1}, which enables periodic collection of node-level KPMs. Once the subscription is accepted, the E2 nodes generate RIC Indication messages containing the requested KPM values. These messages are received and parsed by the xApp through the RIC SDK.

The collected KPMs can be stored and managed within the Near-RT RIC domain. To expose them to other ONDT components and to rApps, the implementation uses a Prometheus-based mechanism. The xApp exports the parsed metrics to Prometheus, using 3GPP/O-RAN naming conventions and including cell and node identifiers as labels. This enables filtering, aggregation, and averaging through the Prometheus API, and allows the same monitoring workflow to support both online observation and subsequent data processing.

Dataset generation is implemented as an optional extension of the monitoring workflow. When the dataset-generation rApp is deployed, it retrieves selected metrics from the Prometheus time-series database and stores them in Minio S3 in CSV format. These datasets can then be used for ML-model training and evaluation using data generated through ONDT experiments. The same KPM xApp and Dataset rApp can also be combined with control-oriented rApps and xApps, allowing metrics to be collected and stored while control algorithms are being validated.

\subsection{Closed-Loop Control}

\begin{figure}
    \centering
    \includegraphics[width=1\linewidth]{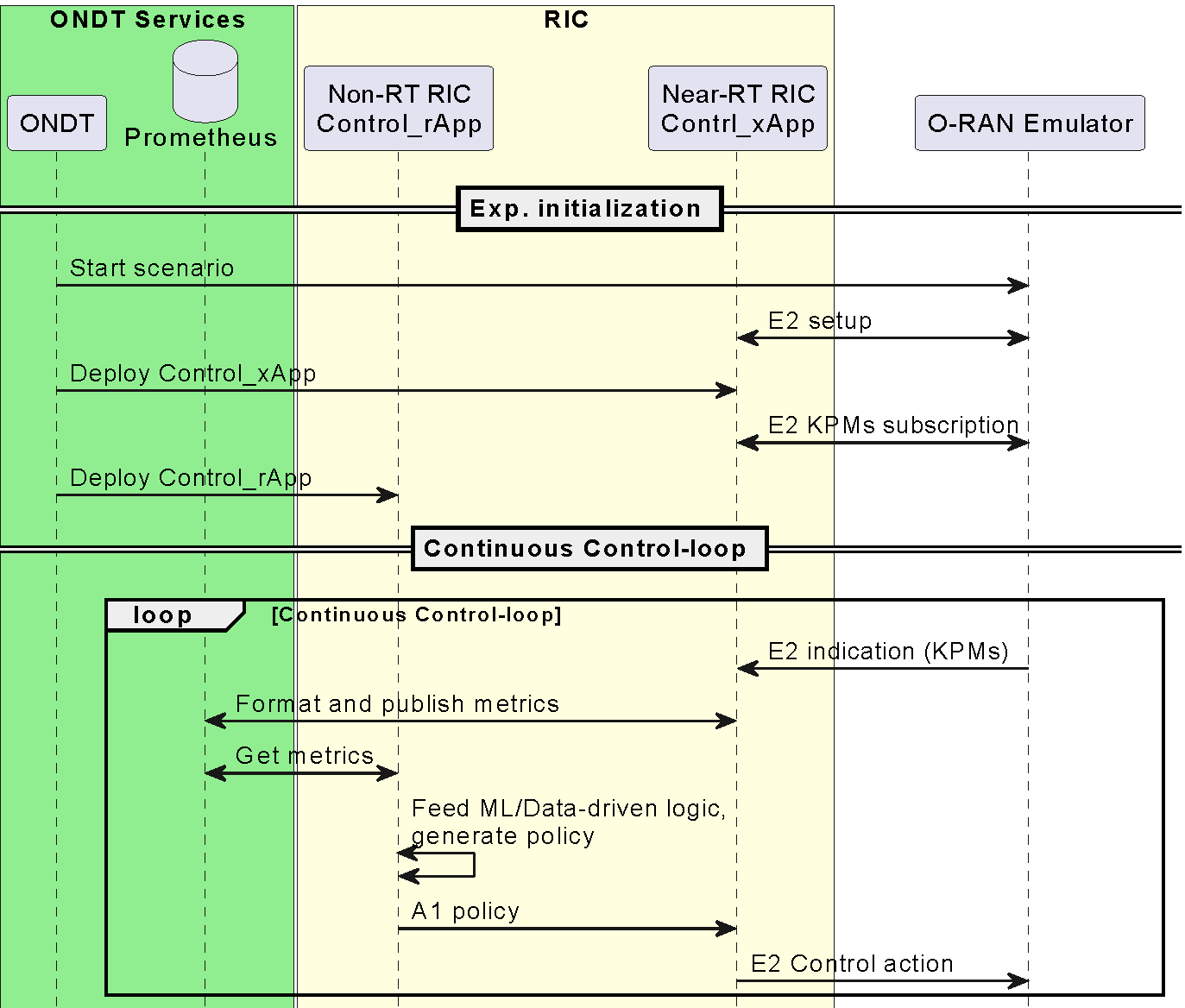}
    \caption{KPM Monitoring and Control workflow.}
    \label{fig:control_workflow}
\end{figure}

The third baseline workflow supported by ONDT is focused on closed-loop control across the Non-RT and Near-RT RIC domains. Figure~\ref{fig:control_workflow} shows the simplified workflow adopted in ONDT. After the selected scenario is started in the O-RAN emulator, ONDT deploys the control xApp in the Near-RT RIC and the control rApp in the Non-RT RIC. Once the E2 setup is completed, the control xApp subscribes to the required KPMs and starts receiving E2 Indication messages from the emulator.

During experiment execution, the control xApp formats the collected measurements and publishes them to Prometheus. The control rApp retrieves the required metrics from Prometheus and applies the corresponding ML- or data-driven control logic to generate the target A1 policy. This policy is then delivered to the control xApp through the A1 framework. After receiving the policy, the control xApp translates it into the corresponding E2 control actions and sends it toward the O-RAN emulator.

This workflow enables ONDT to validate complete cross-RIC control loops under repeatable emulated conditions. In addition, it can be combined with the monitoring and dataset-generation workflow, allowing telemetry collection and control validation to be executed simultaneously within the same experiment.
\section{Functional validation}\label{sec:validation}

The ONDT implementation was validated through a QoS-aware energy-saving use case based on cell ON/OFF control. The objective of the validation presented in this work was not to optimize the control algorithm itself but to verify that the ONDT can execute a complete O-RAN experimentation workflow involving xApp-based KPM collection, rApp-assisted policy generation, A1 policy delivery, and E2-based  control over an emulated RAN scenario.

\subsection{Scenario Generation and Use Case}

The validation scenario is derived from an operational MNO dataset and targets the emulation of representative RAN load patterns. The MNO dataset is processed to extract the load profiles of these target cells. To reproduce the observed day-night traffic pattern, the emulated load is generated by fixing the throughput per UE according to the target QoS level and varying the number of active UEs over time. A parser converts the processed load profiles into a JSON configuration for the Keysight RICtest emulator, which instantiates the corresponding RAN scenario and produces the measurements consumed by the xApps and rApps deployed through ONDT. This process allows ONDT to execute repeatable experiments with realistic traffic evolution, supporting the validation of monitoring, dataset generation, and closed-loop control workflows.

The validation use case considers the selective deactivation of capacity cells during low-load periods, while maintaining traffic service through the remaining coverage infrastructure. The decision logic is constrained by operator-defined QoS requirements, expressed through a target throughput and an outage tolerance. In the validation setup, these parameters are used only to instantiate a representative closed-loop control workflow, while the detailed design and evaluation of the AI/ML model are outside the scope of this paper and can be found in our previous work \cite{practical}.

Each site is represented in the emulator by a 5G capacity cell and associated coverage cells. The 5G cell is modeled as the capacity layer, while the traffic supported by the 4G carriers in the original dataset is aggregated into the corresponding coverage cells of the same site. Capacity cells are modeled with 100~MHz bandwidth, while coverage cells are modeled with 50~MHz bandwidth according to the dataset parameters. For the considered validation, three representative sites with different load levels were selected from the MNO dataset: one highly loaded, one moderately loaded, and one lightly loaded. Each site provides a different energy-saving potential according to the target QoS constraints, defined in this validation as a requested throughput of 5~Mbps and an outage tolerance of 15\%. The selected scenario emulates 72~hours of traffic, with aggregate peaks of 1000 active UEs and 4~Gbps.

\begin{figure}
    \centering
    \includegraphics[width=1\linewidth]{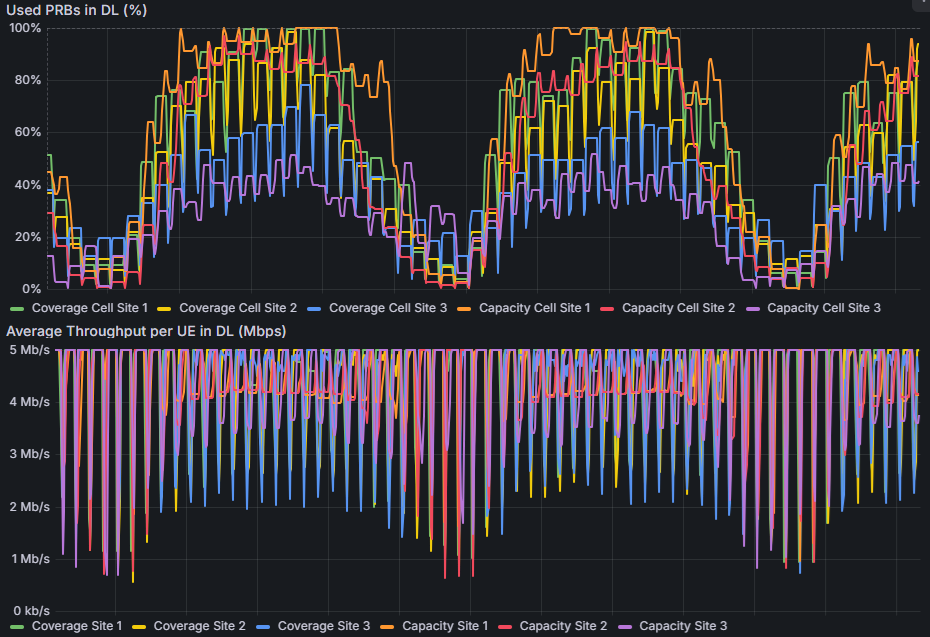}
    \caption{KPMs collected during scenario execution without rApp and xApp. The upper plot shows the downlink PRB usage for coverage and capacity cells across the selected sites, while the lower plot shows the average downlink throughput per UE.}
    \label{fig:results_withoutrapp}
\end{figure}

Figure~\ref{fig:results_withoutrapp} shows an example of the KPMs collected during scenario execution without rApp. The upper plot reports the downlink PRB usage for the selected coverage and capacity cells, while the lower plot reports the average downlink throughput per UE. These measurements are obtained through the E2-KPM workflow and exposed through the ONDT monitoring stack. The PRB usage reflects the different load levels across sites and cells, while the UE throughput plot shows that the emulated scenario maintains the target service level most of the time. Note that the low-throughput peaks observed in the figure are caused by transient simulator effects when modifying the number of active UEs at each step; therefore, they are not considered representative data for the validation.

\subsection{Evaluated control-logic and validation results}

In the implemented validation, the rApp control logic uses the E2-KPM measurements exposed by the xApp to decide the ON/OFF status of the capacity cells. The OFF decision is based on the inference of the ML model embedded in the rApp, which uses recent PRB-usage measurements from the selected sites to determine whether a capacity cell can be switched off during the next control period without degrading the QoS of the flows. The ON decision is based on monitoring the actual UE throughput associated with the capacity cell, also exposed through E2-KPM, and switching the cell back on when the outage constraint is violated. In both cases, stable metric values must be observed over a defined period to avoid instability and ping-pong effects.

The control action is implemented through the generation of O-RAN Energy Saving policies by the rApp, using the A1 \texttt{esObjectives} and \texttt{esPercentage} fields~\cite{oran_a1}. The xApp then maps the received policy into the corresponding E2-CCC message using the \texttt{O-CESManagementFunction} structure~\cite{oran_e2}.

\begin{figure}
    \centering
    \includegraphics[width=1\linewidth]{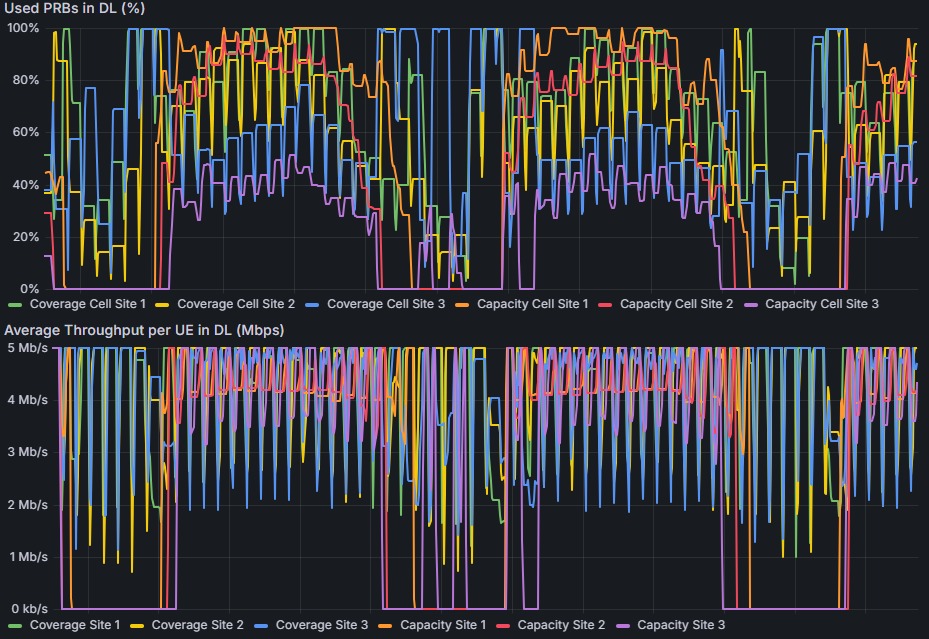}
    \caption{KPMs collected during scenario execution with rApp and xApp. The upper plot shows the downlink PRB usage for coverage and capacity cells across the selected sites, while the lower plot shows the average downlink throughput per UE.}
    \label{fig:results_withrapp}
\end{figure}

Figure~\ref{fig:results_withrapp} summarizes the energy-saving validation results. The figure reports the behavior of the selected cells during the experiment and the corresponding impact of the control decisions. The obtained results yield approximately 35\% average energy savings, at the cost of a 3--4 percentage-point increase in throughput outage.

These results verify that ONDT can execute the complete closed-loop workflow and collect the outputs needed to assess the effect of the control action. This includes telemetry collection through E2-KPM, metric exposure through Prometheus, policy generation in the Non-RT RIC domain, A1 policy delivery, xApp policy processing, and E2-based control toward the emulator. A detailed evaluation of the AI/ML model, the control-logic and of the optimization trade-off between energy saving and QoS is left outside the scope of this paper.

\section{Conclusion}\label{sec:conclusion}
This paper presented ONDT, a modular and cloud-native O-RAN Digital Twin framework developed within SUNRISE-6G for controlled xApp/rApp experimentation. ONDT coordinates the SUNRISE-6G Portal, RIC platforms, and RAN emulation layer to execute repeatable workflows covering rApp/xApp onboarding, scenario execution, KPM monitoring, dataset generation, A1 policy delivery, and E2-based control. The validation, based on an AI-driven QoS-aware energy-saving control loop using Juniper Non-RT and Near-RT RICs and the Keysight RICtest emulator, demonstrated that ONDT can execute complete experimentation workflows over dataset-based RAN scenarios.

Future work will extend ONDT with additional scenarios and southbound adapters, while improving automation for experiment configuration, execution, and result analysis.

\bibliographystyle{unsrt}
\bibliography{references.bib}

@ARTICLE{understanding_oran,
  author={Polese, Michele and Bonati, Leonardo and D’Oro, Salvatore and Basagni, Stefano and Melodia, Tommaso},
  journal={IEEE Communications Surveys \& Tutorials}, 
  title={{Understanding O-RAN: Architecture, Interfaces, Algorithms, Security, and Research Challenges}}, 
  year={2023},
  volume={25},
  number={2},
  pages={1376-1411},
  doi={10.1109/COMST.2023.3239220}}

@article{rimedo,
  author  = {Kryszkiewicz, Paweł and Kułacz, Łukasz and Pakuła, Marcin and Dryjański, Marcin and Hoffmann, Marcin and Skrzypczak, Piotr and Lehmann, Heiko and Stahn, Martin},
  journal = {{IEEE Communications Magazine}},
  title   = {{Large-scale wireless network management via {Open-RAN} Tandem Apps: Cell on/off switching use case}},
  year    = {2026},
  note    = {To appear (Early Access)},
url="https://arxiv.org/abs/2604.24465",
}

@techreport{orandt,
  author      = {{O-RAN Alliance}},
  title       = {{Research Report on Digital Twin RAN Use Cases}},
  institution = {O-RAN next Generation Research Group (nGRG)},
  type        = {Research Report},
  number      = {O-RAN.nGRG.Research-Report-DT-v01.00},
  year        = {2024},
  url         = {https://www.o-ran.org/specifications}
}

@ARTICLE{colosseum,
  author={Polese, Michele and Bonati, Leonardo and D'Oro, Salvatore and Johari, Pedram and \textit{et al.}},
  journal={IEEE Open Journal of the Communications Society}, 
  title={{Colosseum: The Open RAN Digital Twin}}, 
  year={2024},
  volume={5},
  number={},
  pages={5452-5466},
  doi={10.1109/OJCOMS.2024.3447472}}

@INPROCEEDINGS{dt_eucnc,
  author={Zafar, Hammad and Tohidi, Ehsan and Kasparick, Martin and Stańczak, Sławomir},
  booktitle={2025 Joint European Conference on Networks and Communications \& 6G Summit (EuCNC/6G Summit)}, 
  title={{An Open Source Digital Twin Framework for O-RAN}}, 
  year={2025},
  volume={},
  number={},
  pages={286-291},
  doi={10.1109/EuCNC/6GSummit63408.2025.11036878}}

@ARTICLE{dt_commag,
  author={Nguyen, Huan X. and Sun, Kexuan and To, Duc and Vien, Quoc-Tuan and Le, Tuan Anh},
  journal={IEEE Communications Magazine}, 
  title={{Digital Twin for O-RAN Toward 6G}}, 
  year={2025},
  volume={63},
  number={3},
  pages={174-181},
  doi={10.1109/MCOM.003.2400016}}

@inproceedings{twinning,
author = {Bonati, Leonardo and Shirkhani, Ravis and Fiandrino, Claudio and Maxenti, Stefano and \textit{et al.}},
title = {{Twinning Commercial Network Traces on Experimental Open RAN Platforms}},
year = {2024},
isbn = {9798400704895},
publisher = {Association for Computing Machinery},
address = {New York, NY, USA},
url = {https://doi.org/10.1145/3636534.3697320},
doi = {10.1145/3636534.3697320},
booktitle = {Proceedings of the 30th Annual International Conference on Mobile Computing and Networking},
pages = {1914–1921},
numpages = {8},
location = {Washington D.C., DC, USA},
series = {ACM MobiCom '24}
}

@ARTICLE{dataset_nec,
  author={Salvat, J. Xavier and Ayala-Romero, Jose A. and Zanzi, Lanfranco and Garcia-Saavedra, Andres and Costa-Perez, Xavier},
  journal={IEEE Communications Magazine}, 
  title={Open Radio Access Networks (O-RAN) Experimentation Platform: Design and Datasets}, 
  year={2023},
  volume={61},
  number={9},
  pages={138-144},
  doi={10.1109/MCOM.003.2200711}}

@misc{EricssonEIAP2026,
  author       = {{Ericsson}},
  title        = {Intelligent Automation Platform},
  howpublished = {\url{https://www.ericsson.com/en/ran/intelligent-ran-automation/intelligent-automation-platform/ecosystem}},
  year         = {2026},
  note         = {Accessed: 2026-06-09}
}

@misc{NokiaMantaRaySMO2026,
  author       = {{Nokia}},
  title        = {MantaRay {SMO}: Service Management and Orchestration},
  howpublished = {\url{https://www.nokia.com/mobile-networks/ran-operations/mantaray-smo/}},
  year         = {2026},
  note         = {Accessed: 2026-06-09}
}

@misc{telcofabric,
  author       = {{BubbleRAN}},
  title        = {{TelcoFabric}: Transforming Telecom Infrastructure},
  howpublished = {\url{https://bubbleran.com/news/telcofabric/}},
  year         = {2026},
  note         = {Accessed: 2026-06-09}
}

@INPROCEEDINGS{testbed,
  author={Navarro, Juan and Castro, Alberto and Rattaro, Claudina},
  booktitle={2025 13th Wireless Days Conference (WD)}, 
  title={O-RAN Testbeds: From Experimental Deployments to Simulation Frameworks}, 
  year={2025},
  volume={},
  number={},
  pages={1-7},
  doi={10.1109/WD67713.2025.11302581}}

@ARTICLE{interoperability,
  author={Mehran, Farhad and Turyagyenda, Charles and Kaleshi, Dritan},
  journal={IEEE Access}, 
  title={Experimental Evaluation of Multi-Vendor 5G Open RANs: Promises, Challenges, and Lessons Learned}, 
  year={2024},
  volume={12},
  number={},
  pages={152241-152261},
  doi={10.1109/ACCESS.2024.3476963}}

@INPROCEEDINGS{practical,
      title={{A Practical AI-Driven Strategy for Cell On/Off Switching under Adaptable QoS Constraints}}, 
      author={David Reiss and Miguel Catalan-Cid and Daniel Camps and Oriol Sallent},
      year={2026},
      booktitle={IEEE International Conference on Communications (ICC'2026)}
 
}

@techreport{oran_a1,
  author       = {{O-RAN Alliance}},
  title        = {{O-RAN A1 interface: Type Definitions 11.0}},
  institution  = "{O-RAN Alliance Working Group 2}",
  number       = "{O-RAN.WG2.TS.A1TD-R005-v11.0}",
  year         = "2025",
  month        = "October"
}

@techreport{oran_e2,
  author       = {{O-RAN Alliance}},
  title       = {{O-RAN E2 Service Model (E2SM) Cell Configuration and Control 5.0}},
  institution  = "{O-RAN Alliance Working Group 3}",
  number      = {{O-RAN.WG3.TS.E2SM-CCC-R004-v05.00}},
  month       = feb,
  year        = {2025}
}
\end{document}